# Classification of Skin Cancer Images using Convolutional Neural Networks

**Kartikeya Agarwal**
Student of Computer
Science and Engineering
Netaji Subhas University of Technology
kartikeya.co19@nsut.ac.in

**Tismeet Singh**
Student of Computer
Science and Engineering
Netaji Subhas University of Technology
tismeet.co19@nsut.ac.in

## Abstract

Skin cancer is the most common human malignancy(American Cancer Society) which is primarily diagnosed visually, starting with an initial clinical screening and followed potentially by dermoscopic(related to skin) analysis, a biopsy and histopathological examination. Skin cancer occurs when errors (mutations) occur in the DNA of skin cells. The mutations cause the cells to grow out of control and form a mass of cancer cells. The aim of this study was to try to classify images of skin lesions with the help of convolutional neural networks. The deep neural networks show humongous potential for image classification while taking into account the large variability exhibited by the environment. Here we trained images based on the pixel values and classified them on the basis of disease labels. The dataset was acquired from an Open Source Kaggle Repository(Kaggle Dataset)which itself was acquired from ISIC(International Skin Imaging Collaboration) Archive. The training was performed on multiple models accompanied with Transfer Learning. The highest model accuracy achieved was over 86.65%. The dataset used is publicly available to ensure credibility and reproducibility of the aforementioned result.

## 1 Introduction

The skin is the largest organ of the body which protects all the internal organs from the harsh world outside. It helps regulate temperature and protects against infections. The skin consists of three layers:

- Epidermis
- Dermis
- Hypodermis

Cancer can be traced when a mass of healthy skin cells changes their property and start growing uncontrollably forming a mass called tumour. This tumour can be a benign growth, meaning that there is no harm to life and it cannot grow or spread to other areas whereas it can be malignant which means that it is interrupting the natural flow of things and can grow as well as spread to other parts of the body.

Skin cancer is the most common form of cancer generally associated with the over-exposure to sun or other forms of high energy particles which destroy the genetic build-up of the cell.

As the world steps out into a new millennium, the amount of pollution in the air is growing at an exponential rate paving a way for skin diseases like skin cancer to thrive. The Ozone Layer is the only protection from the harmful ultraviolet rays of the sun, which is the largest most probable cause for skin cancer. Due to the growing levels of air pollution, this layer is depleting geometrically and thus the number of skin cancer cases are on the rise. Skin cancer can be broadly classified into two major categories: Melanoma (Malignant) and non-melanoma (Benign). Melanoma is one of the deadliest kinds of cancer. However, the detection of this cancer at an early stage can help in improving the chances of survival.

Various methods were considered and studied before the selection of the actual functions for this thesis. The best and the most advanced of the available machine learning resources were employed.

Skin cancer detection through histopathology can be achieved by augmenting the power of deep learning. The classification of skin cancer lesions through specialized convolutional neural networks can help the medical community in detecting and possibly curing the disease at the early stages. Some early stipulations related to this cancer are discolouration or inflammation of the skin, itching and bleeding of skin patches, moles or red, waxy bumps on the skin due to the cancerous cells.

The dataset was extracted from an open Source Kaggle repository which itself was a part of the ISIC (International Skin Image Collaboration)(Kaggle Dataset). The images were extracted and were split into test and train segments. An adequate level of image augmentation was applied to expand the dataset with modified versions of images. The test set contained 350 images while the training dataset contained over 2900 images comprising two distinct classes, namely, “benign” and “malignant”.

Within the scope of this research, we found that classification using K-nearest Neighbours or Support Vector Machines or even Decision Trees yielded low precision as well as accuracies. On further studying the mathematics behind classification, it came into light that the most sophisticated method to achieve the desired results was to use Deep Learning models. We tried various mathematical models with and without the use of Transfer Learning, but in the end, it was concluded that the depth and quality of activation provided by pretrained models, were of no-match, hence we paired our knowledge of mathematics and created a Dense Convolutional Network model which gave us accuracy of over 86.6%.

Deep Learning, a robust and powerful protégé of Artificial Intelligence, provided us with tools to perform effective and accurate image classification. Its structure and operation resembled the human brain with neurons firing across the brain, sending information, classifying data, drawing inferences and producing results. The structure used by Deep Learning consists of a stack of layers collectively known as a Neural Network. As the name suggests neural networks work like neurons in a brain like identifying patterns and forming predictions.

The aim of the study is to try to get a reasonably faster and more accessible method of detection of skin cancer. The earlier it is detected, the higher is the chance that the sufferer will recover. According to the American Academy of Dermatology Association, if found early, skin cancer is highly treatable. The model trained is just a precursor to a skin biopsy. The only way to know for sure, the presence of skin cancer is to visit a Dermatologist and get a skin biopsy.

## 1.1 Related Works

To keep the paper as relevant as possible a lot of other papers and journals were referred and consulted. The author (Mohammad Ashraf Ottam) recorded a maximum accuracy of 0.74 with his application of mathematical models on an augmented dataset of 3000 images. The author in his paper too aspires to promote the early diagnosis of melanoma via further development of machine learning models.
Further a few more papers (Giotis, I., Molders, N., Land, S., Biehl, M., Jonkman, M. F., & Petkov, N. (2015)) were consulted, here the authors concluded an 81% accuracy, his proposed system made use of non-dermoscopic images of lesions from which he extracted the lesion regions and computed the results based on the colour and texture.

# 2 Materials and Methods

## 2.1 Dataset

A dataset of 2947 histopathology images were considered for this study.
The dataset was acquired from an open source Kaggle Repository, a subset of the ISIC (International Skin Imaging Collaboration) archive. The dataset consists of two classes namely, benign and malignant. The test dataset comprises 350 images taken from the Kaggle repository. The dataset chosen is publicly distributed and open source to ensure better credibility of the models.

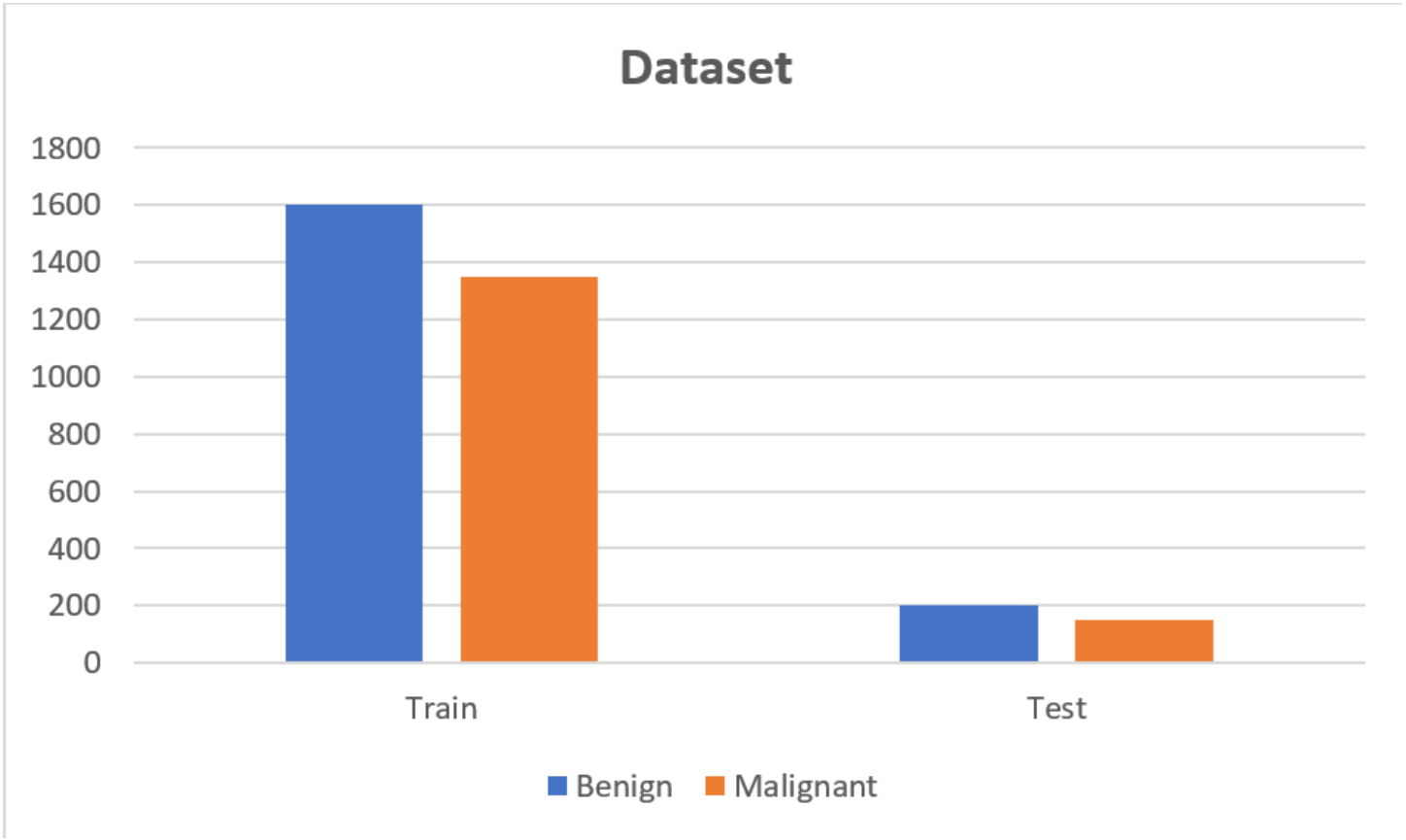

**Fig 1: Distribution of dataset**

The division among the 2 classes were as follows:

- For training and validation purpose
    1. 1600 Benign skin mole images
    2. 1347 Malignant skin mole images
- For testing purpose
    1. 200 Benign skin mole images
    2. 150 Malignant skin mole images

All the images were resized to 176 x 176 for uniformity and improved testing metrics.

## 2.2 Methodology

The aim of the study is to determine the efficiency and the accuracy of different models in-order to classify images between being benign and malignant. The study takes into account multiple Transfer Learning models paired with a set of Dense Layers to predict the class of the image.

## 2.3 Augmentation

The Dataset obtained was adequately augmented. Data Augmentation is a technique that can be used to exaggerate the dataset to introduce variability and expand the data so that a more robust and reliable model can be trained. The augmentation parameters we selected extremely carefully so as not to disturb the concerned image parameters that factor in deciding the outcomes. Augmentation included:

- The images were re-scaled to a re-scaling ratio of 1:255.0.
- The images underwent a shear of 0.2 degrees, which is the shear angle, anticlockwise.
- The images were randomly flipped horizontally as well as vertically so as to preserve the lesion marks and create the new synthetic data.
- A zoom augmentation introduced, randomly zooms the image in and either adds new pixel values around the image or interpolates pixel values.

All the images generated were passed through a generator and were passed into the model as a whole with a height and width of (176,176). All the images were passed through all three of the channels, i.e., red, green and blue (R, G, B).

## 2.4 Convolutional Neural Network

Convolutional Neural Network is a class of neural networks applied in deep learning mostly to analyse image or visual data. A Convolutional Neuron is made up of learnable weights and biases which are tuned to generate desired results. Thousands and millions of these neurons when arranged in a specific order with other neurons result in a Convolutional Neural Network or CNN. CNNs are mainly applied to classify and order image data and cluster them together if they appear analogous, and then perform object recognition.

Data or image is convolved using filters or kernels. Filters are small matrices that we engage across the dataset with the help of a sliding window. The depth of the image is the same as the input, for a coloured RBG image value of depth is 4, a filter of depth 4 would also be applied to it.

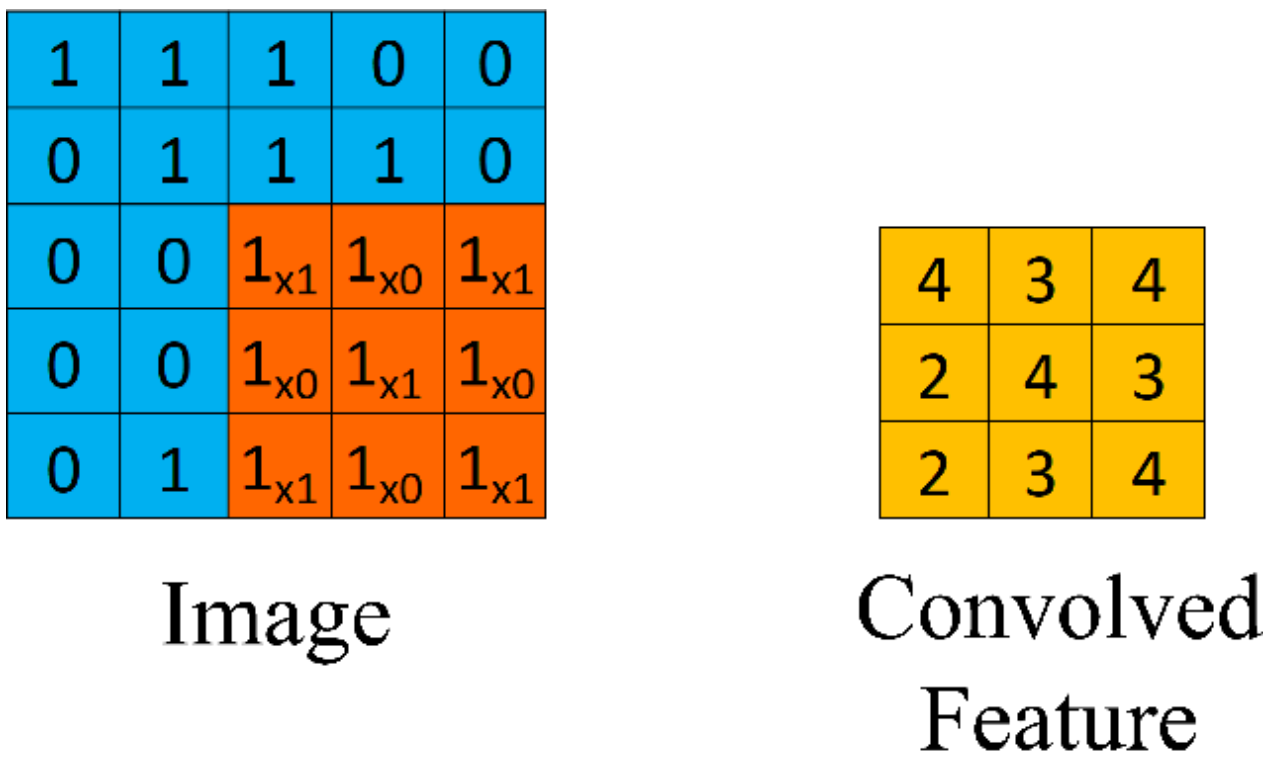

**Fig 2: Visual representation of Convolution function**

The next part is, we apply a rectifier function (generally Rectified Linear Unit) to increase the non-linearity of the Network. And finally, a pooling layer is used to down sample the features, which is applied through the volume of the image. We finally use a fully connected Layer which flattens the entire 3d feature matrix into a single column which is then fed further to the neural network for processing. All the Layers; Convolutional Layer, Activation Layers, Pooling Layers, and Fully Connected Layers are interconnected as previously mentioned to form a Convolutional Neural Network.

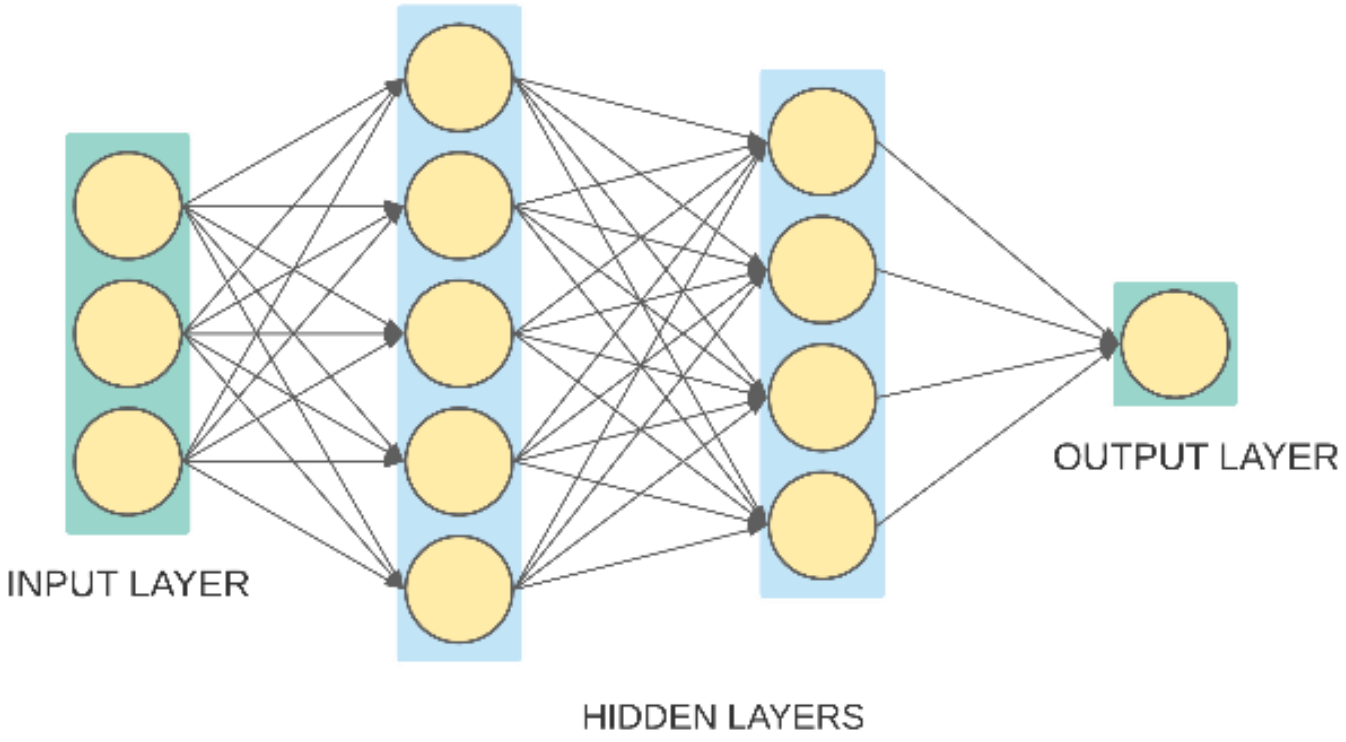

**Fig 3: Visualization of a Neural Network**

## 2.5 ReLU

The most frequently utilized method of activation function in deep learning algorithms is the Rectified Linear Unit. The function returns 0 if it receives any negative input, but for any positive value x it returns that value back. So, it can be written as:

$$f(x) = \max(0, x)$$

## 2.6 Leaky ReLU

The definition of the Leaky ReLU activation function is instead of using 0 for negative values of inputs(x), we define it as an extremely small linear component of x. It helps by-passing the dead neuron problem caused by Rectified Linear activation. Leaky ReLU is a better activation function as compared to sigmoid and tanh because it is less prone to saturation and has better sensitivity as compared to other activation functions. Leaky ReLU looks like a linear activation function but is actually a nonlinear function thereby helping in defining complex associations in the data

$$f(x) = \max(a, x), \text{where a is a very small value}$$

## 2.7 Batch Normalisation

Normalization is a technique of changing the variables in a particular type of data to a specific scale without affecting the contrast or shape of the image. Batch Normalization refers to a procedure of normalization of the effect to a layer for every batch. It helps in reducing the amount of computational power and number of training epochs required for a deep learning model. When a deep neural network is trained the inputs of layers may change when the weights are updated per mini batch. This can cause internal covariate shift, that is changes in the distribution of inputs to a layer for each mini batch.

Batch Normalization for a neural network means scaling the output of the layer by normalizing the activations of each input variable per mini batch. Batch Normalization helps to make the model more robust and less prone to effects caused by hyperparameter tuning.

## 2.8 Flatten

Flatten is a technique adopted in the conversion of data to a 1D array as an input for the next layer. The output vector or the feature matrix map is flattened to get a single feature vector which can be used as an input to the next layer to be processed further.

## 2.9 Pooling

Pooling is one of the critical layers in a deep neural network. The pooling layer is responsible for the reduction of dimensionality or spatial size of a feature matrix. The pooling layer helps in reducing the computational power required to train a neural network. It is also critical in the extraction of dominant features in the data.

There are two major types of pooling layers which can be incorporated in a deep neural network: Max Pooling and Average Pooling.

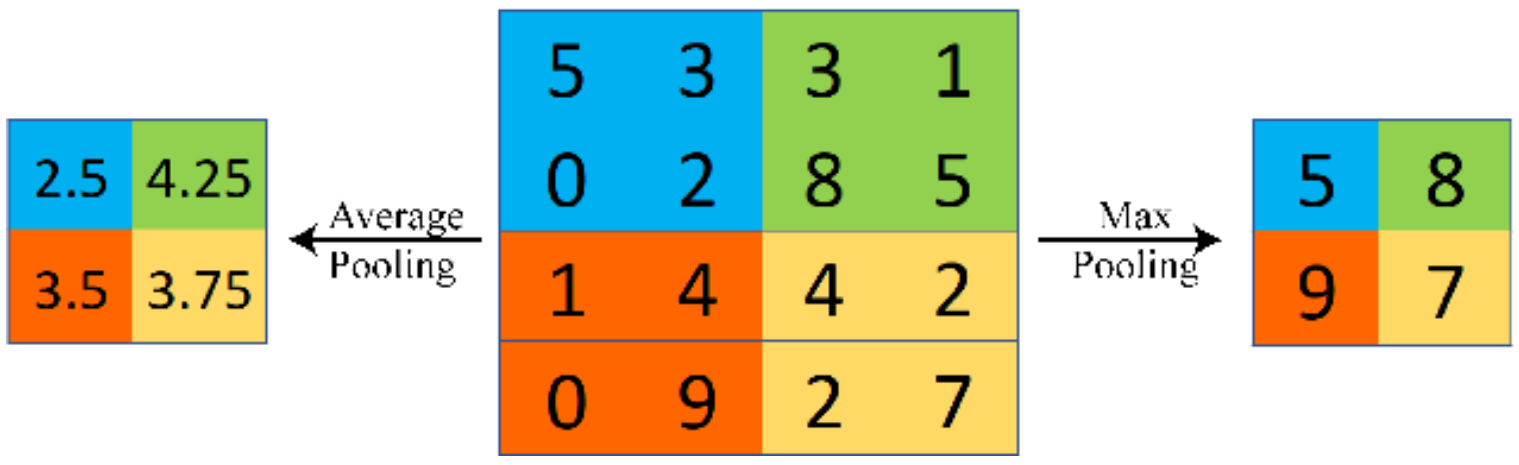

**Fig 4: Average Pooling vs Max Pooling**

Max Pooling, as the name suggests, gives the maximum value for the image area covered by the kernel. Similarly, Average Pooling returns the average of the values for the image area covered by the kernel. In our study we have used Max Pooling instead of Average Pooling as it helps in the elimination of noise or noisy features along with dimensionality reduction whereas Average Pooling only helps in dimensionality reduction. Pooling layer, thus, helps in picking up meaningful information from a sea of data which is vital for helping a model learn the features in a dataset. Elimination of noise and provision of only useful features helps in preventing overfitting and speeds up the computation.

## 2.10 Dropout

Dropout Layers are used to decrease the level of overfitting in a large Network. Simply put, a dropout layer will ignore a specified number of random Neurons while training so as to prevent overfitting and bias in the output of the model.

## 2.11 Categorical Crossentropy

Categorical Cross Entropy is a loss function employed in a Network where there is a multiclass classification.

$$Loss = -\sum_{i=1}^{\substack{output \\ size}} (y_i \cdot log\hat{y}_i)$$

This function is an excellent method of differentiating between different classes or discrete probability distributions.

## 2.12 Optimizer and Metrics

Optimizers help in tuning the weights such that the loss function is minimized. Their objective is reducing losses and providing better results. The 'adam' optimizer was used for training the model with a variable learning rate. The learning rate was initialised with a value of 0.0001 but was modified or more accurately, reduced after the validation-loss crossed a specific patience level with a reduction factor of 10^-1. The metric used for the model was 'accuracy' and Early stopping was employed to prevent overfitting or overtraining of the model. Here too, the monitored hyper-parameter was the validation-loss with a high yield patience level.

Activation Functions are generally used to normalize the output of a network to a probability distribution over the output classes. The activation function considered for the model's feature extraction layer was SoftMax.

$$\sigma(\vec{z})_i = \frac{e^{z_i}}{\sum_{j=1}^{K} e^{z_j}}$$

$\sigma$ = softmax

$\vec{z}$ = input vector

$e^{z_i}$ = standard exponential function for input vector

$K$ = number of classes in the multi-class classifier

$e^{z_j}$ = standard exponential function for output vector

$e^{z_j}$ = standard exponential function for output vector

## 2.13 Models Used

The study required meticulous comparison among a plethora of transfer learning models. Several transfer learning models were considered for this study namely DenseNet, XceptionNet, ResNet and MobileNet. DenseNet201 was chosen because of its strong gradient flow, better back propagation and smooth decision boundaries. Xception Net encompasses modified depth wise separable convolution and has the same number of parameters as InceptionV3 but greater computational efficiency. Resnet50 illustrates the concept of skip connection which eliminates the problem of vanishing gradients (problem encountered while training very deep networks) and helps the model to learn the identity function properly for better results. MobileNetV2 also consists of depth wise separable convolution which provides reduced complexity and size.

## 2.14 Gradient Class Activation Maps

Grad CAM is a technique for analysis of the Activations generated by the trained model. Grad CAM simply is a heat-map generated by the Convolutional Layers depicting the importance of a certain feature or the part of an image. It encapsulates the area or region of interest (ROI).

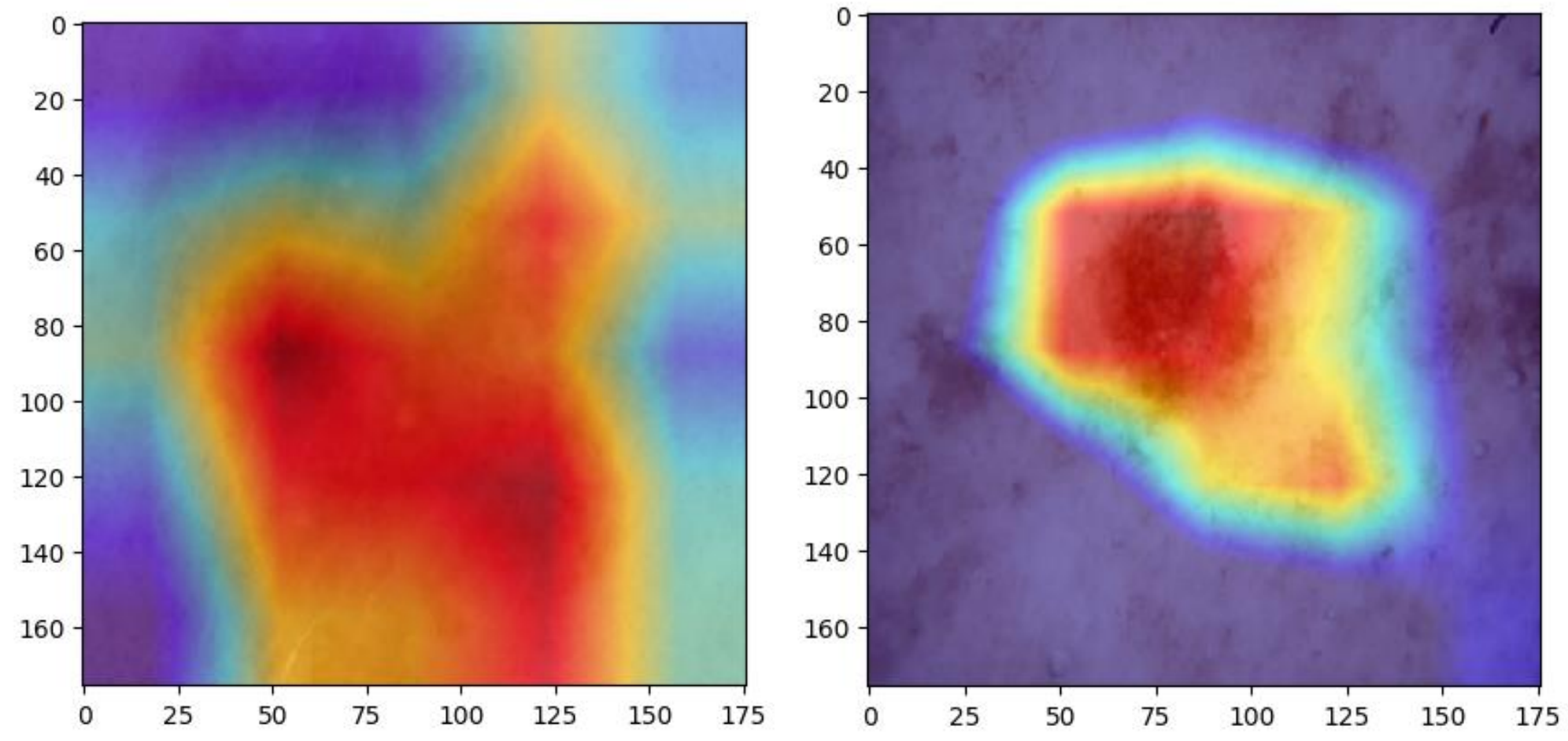

**Fig 5: Gradient Class Activation Maps**

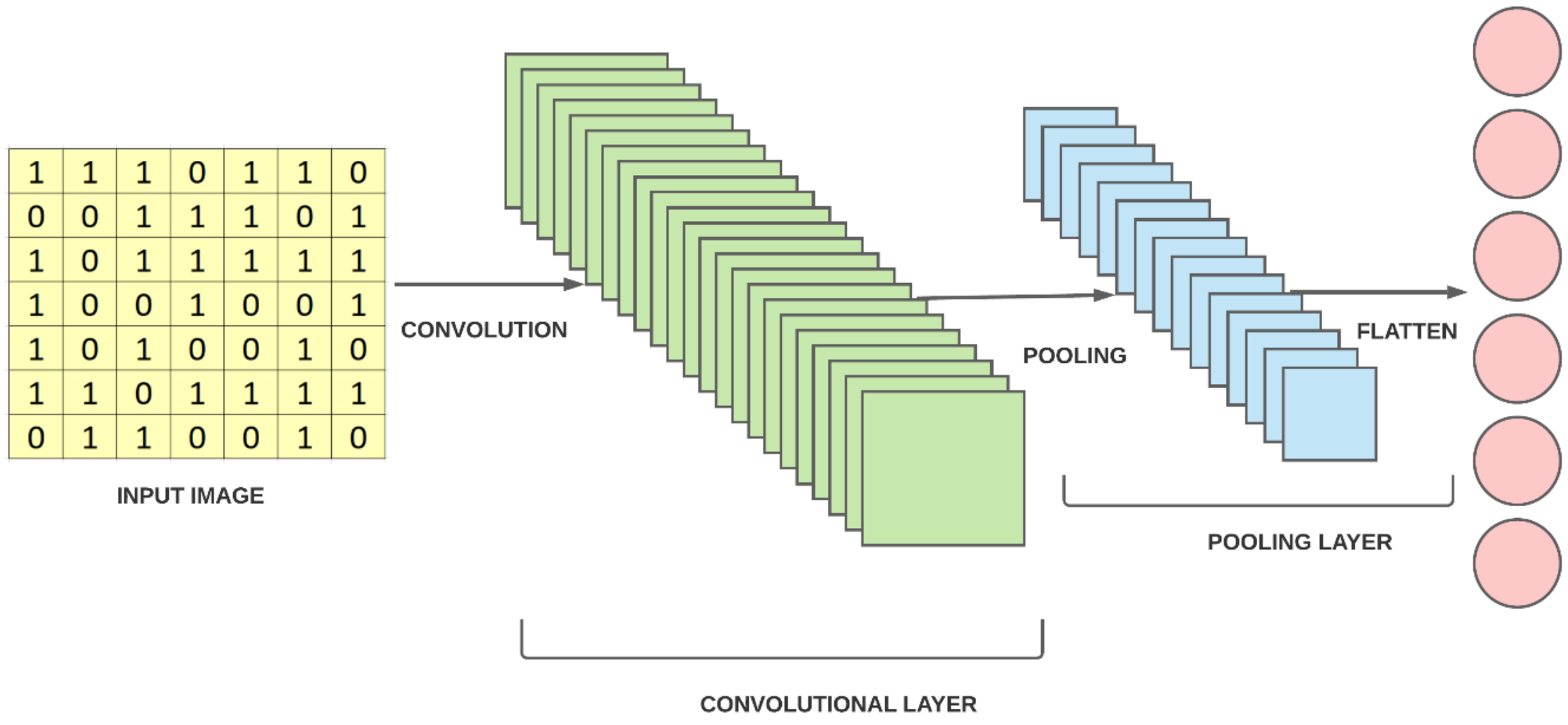

**Fig 6: Structure of a Neural Network**

## 2.15 Training

The Model was trained on the previously mentioned dataset after performing a training and validation split. The aforementioned models were trained with 60 epochs though as mentioned earlier, early-stopping was employed on account of the validation-loss. The steps per epoch were decided as the batch-size divided by the total number of images in the database. Different models yielded different results which are plotted below. The early-stopping was employed to prevent over training and it monitored validation loss as a parameter and at every point, when the model performed worse, the previous weights were restored. We also employed a conditional reduction in learning rate by 1/10 so that whenever the rate of convergence of the function plateaued, the learning rate would be reduced for a more precise reduction in the slope of the hyper-plane.

## 3 Results

The models were tested on a test split of 350 images being 150 from malignant class and 200 from the benign class. The model accuracy, precession, recall and F1-score was calculated for all the models for better insights of their efficiency and operability. The confusion matrix was plotted for each model based on the test data and the metrics were calculated according to the undermentioned formulas.

$$Precision = \frac{True\ Positive}{True\ Postive + False\ Positive}$$

$$Recall = \frac{True\ Positive}{True\ Postive + False\ Negative}$$

$$Accuracy = \frac{True\ Positive + True\ Negative}{True\ Postive + False\ Negative + False\ Positive + True\ Negative}$$

$$F1Score = \frac{2 \times Recall\ \times Precision}{Recall + Precision}$$

## 3.1 DenseNet201

DenseNet201 performed extremely well on the Validation split as well as on the final test split. The Neural Network paired with Dense Layers which in turn complemented with Leaky ReLU gave high accuracies as well as F1 Score. DenseNet helped solve the problem of vanishing-gradients and enabled feature reuse and propagation leading to substantial results and accuracy. For visual insights, the Confusion Matrix derived from the test set and the graphical representation of the metrics used during training and validation steps are shown below.

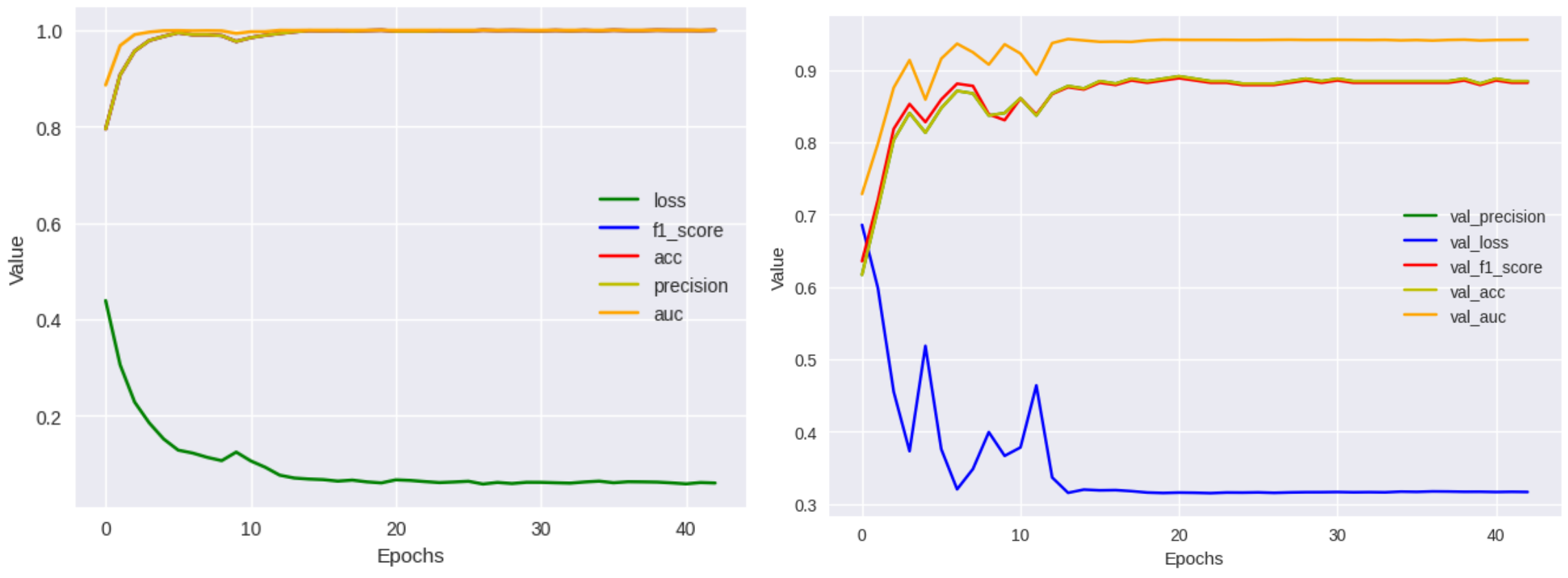

**Fig 7: Graphical Representation of Training and Validation Metrics - DenseNet201**

The accuracy recorded over the test set for this model was 86%.

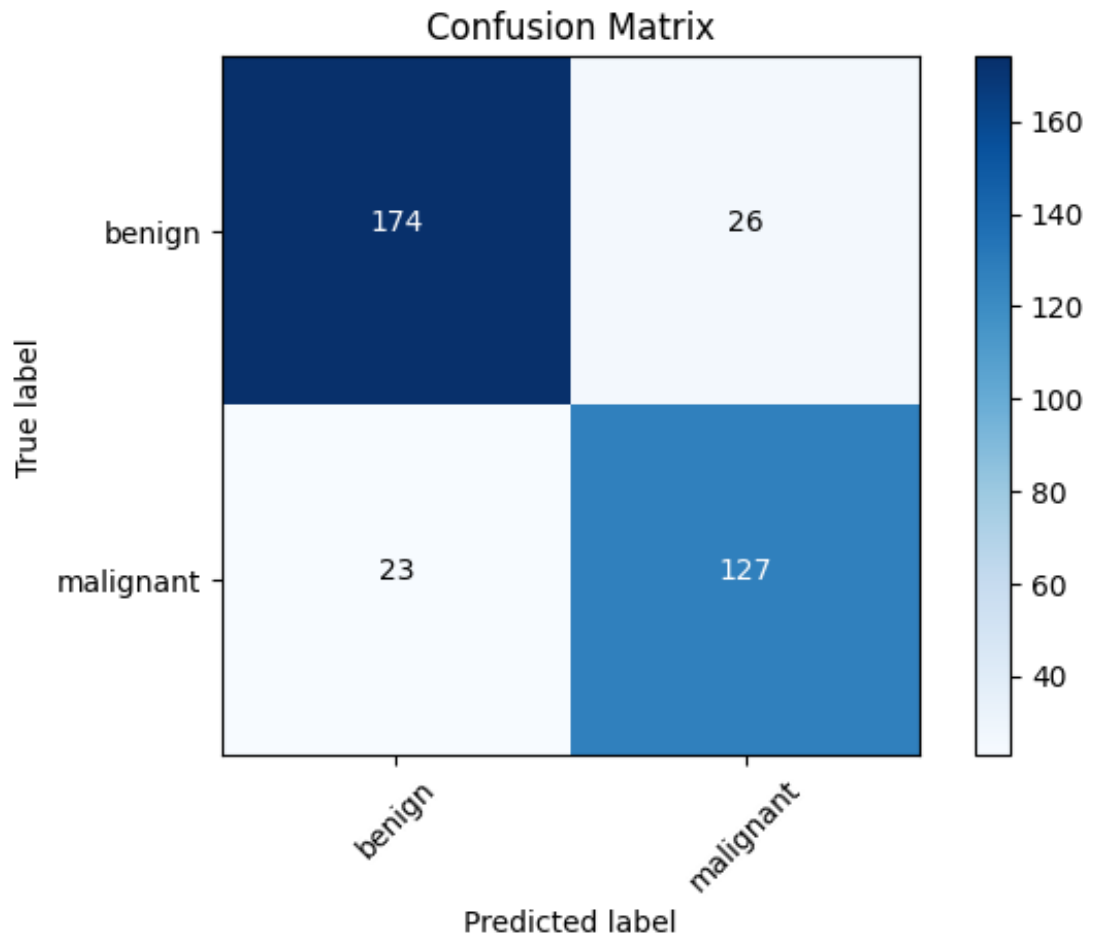

**Fig 8: Confusion Matrix - DenseNet201**

The class activation map for this model was prepared where the activation provided by the last layer is represented as an overlayed heat map. The red-yellow colour depicts the region of interest while the violet-blue colour represents are of low activation.

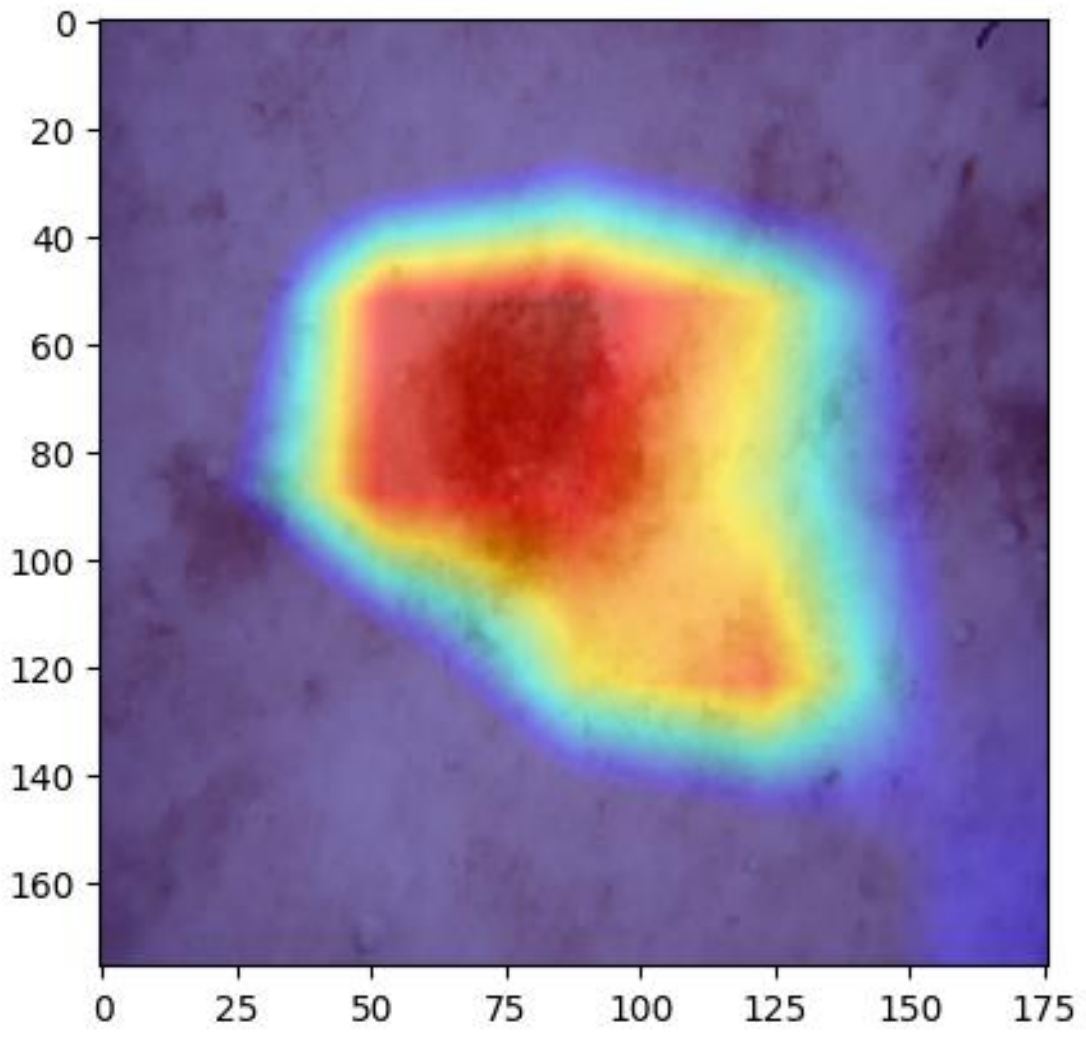

**Fig 9: Grad-CAM - DenseNet201**

## 3.2 ResNet50

The ResNet50 architecture is equipped with skip connections i.e., the input to a layer can be directly passed to some other layer thereby improving the performance of this neural network over others. Along with that, it also helps in solving the problem of vanishing gradients through identity mapping. The confusion matrix along with the metrics used during the training and validations steps were plotted to draw comparisons with the other models used in this study. The model gave an accuracy of over 86.57% on the test data.

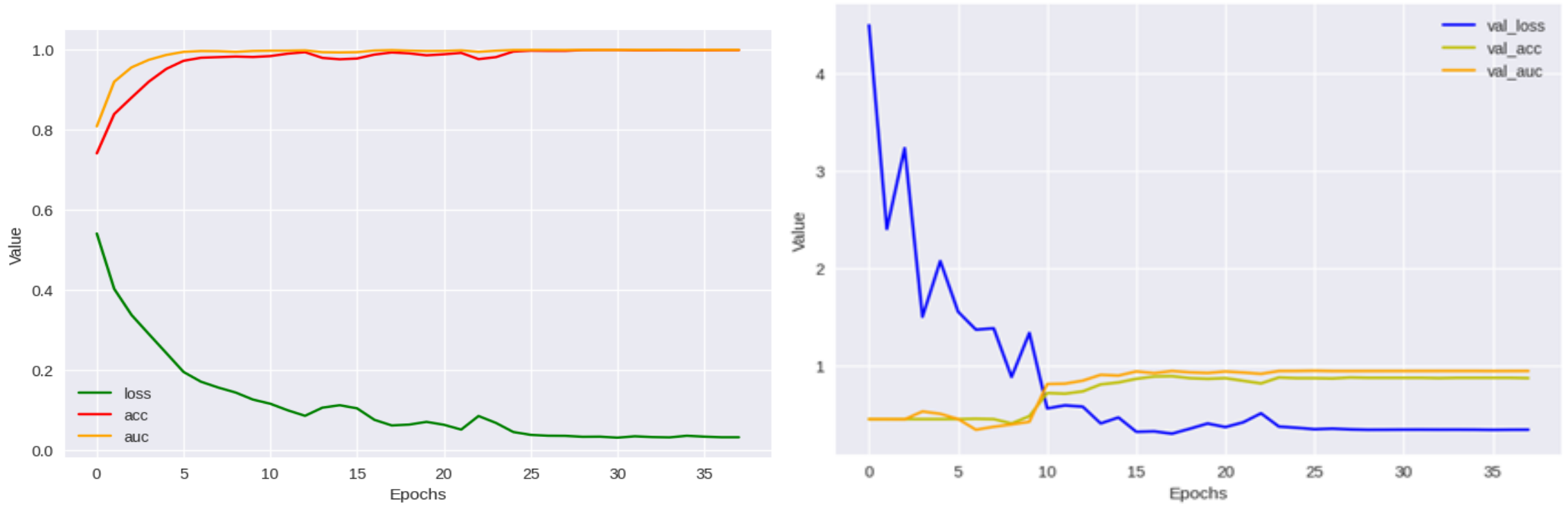

**Fig 10: Graphical Representation of Training and Validation Metrics - ResNet50**

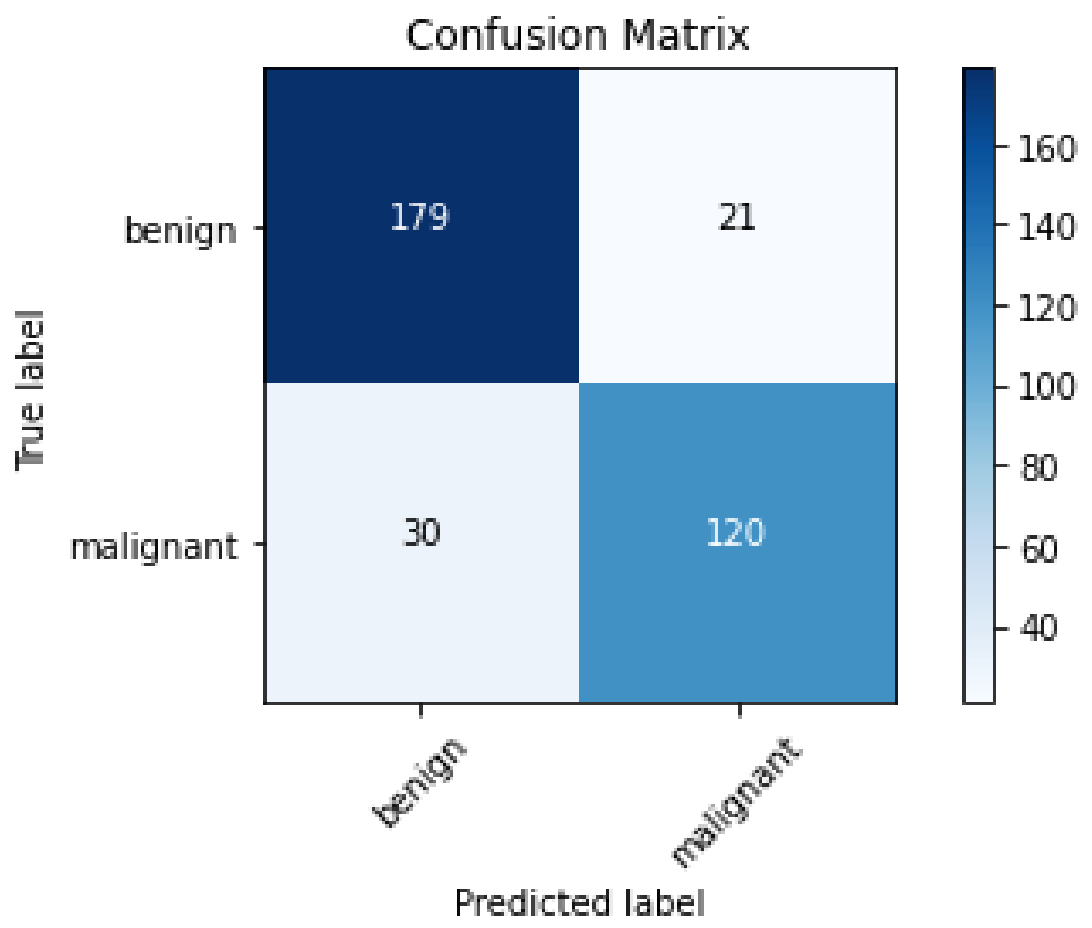

**Fig 11: Confusion Matrix- ResNet50**

The class activation map for this model was prepared where the activation provided by the last layer is represented as an overlayed heat map. The red-yellow colour depicts the region of interest while the violet-blue colour represents are of low activation.

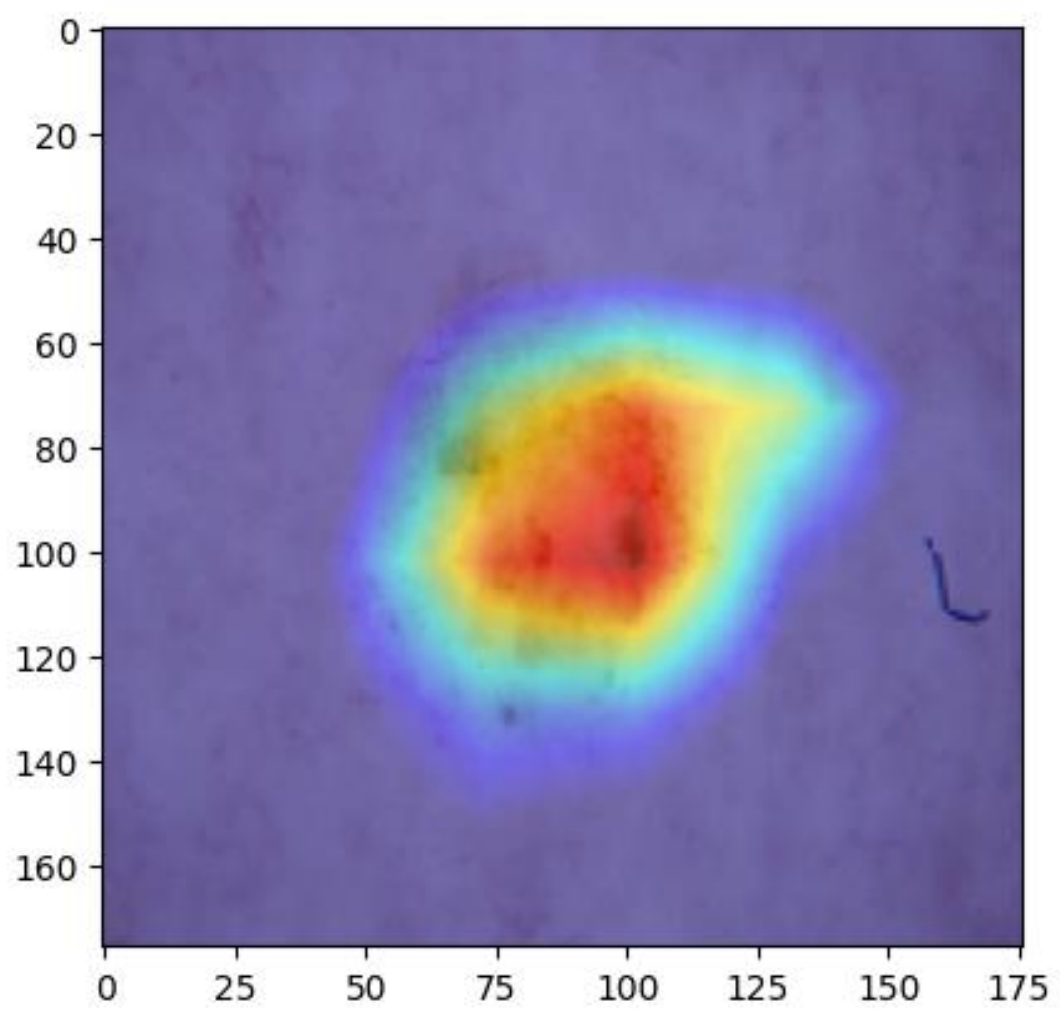

**Fig 12: Grad-CAM - ResNet50**

## 3.3 XceptionNet

The Xception-Net architecture has 36 convolutional layers forming the feature extraction base of the network. The 36 convolutional layers are compiled into 14 different modules in linear geometry. This arrangement helped us gain immense accuracy on the succeeding organization of the dense and batch normalisation layers.

The model was trained on the same set of training dataset and the performance of the Model was analysed consistently with the same set of validation images. The validation graph and training graph were plotted and the confusion matrix was analysed for the test dataset to get a better understanding of the results and for the analysis of the performance of the model. The model gave an accuracy of over 82.5% as is depicted by the charts below.

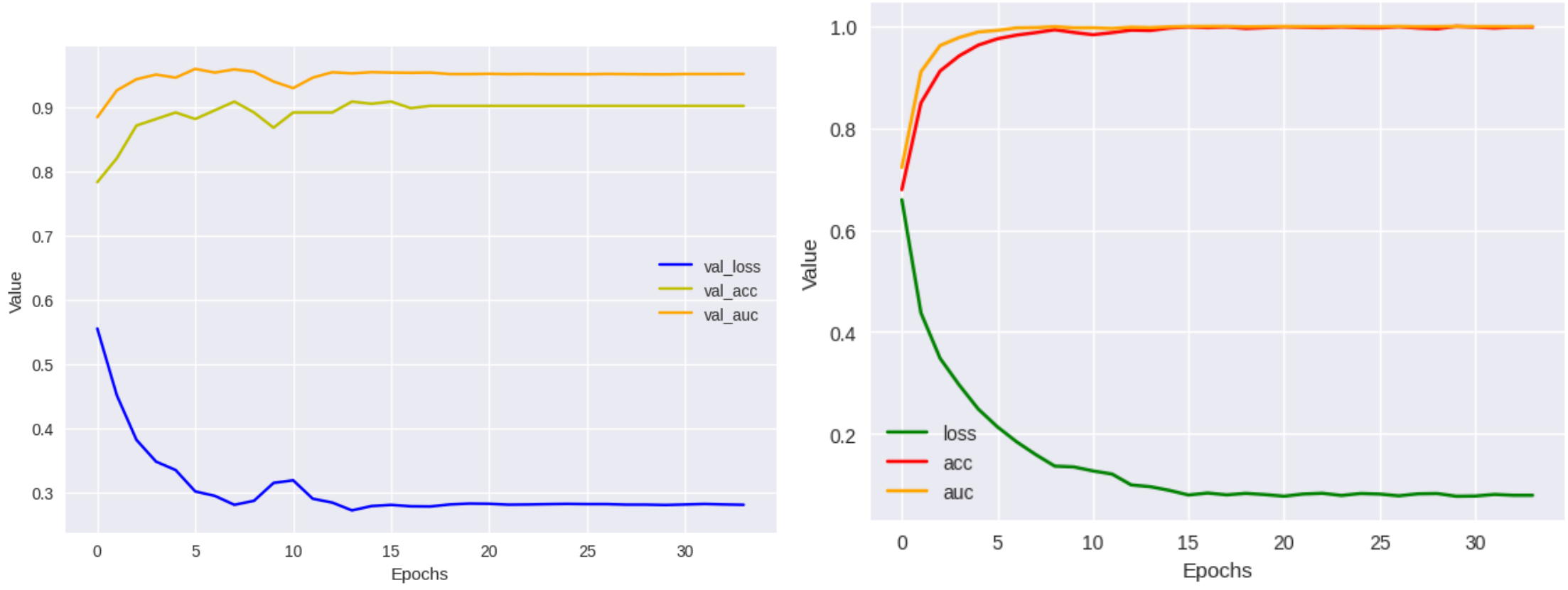

**Fig 13: Graphical Representation of Training and Validation Metrics - XceptionNet**

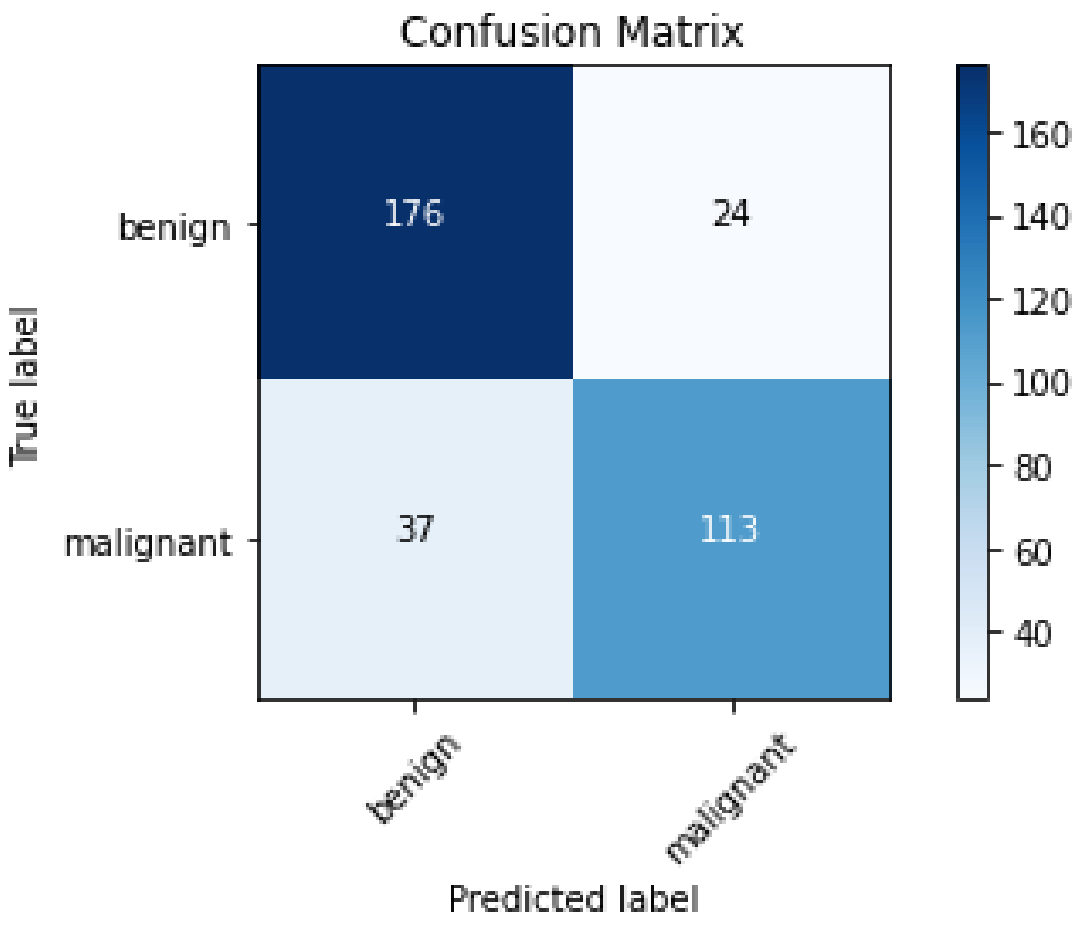

**Fig 14: Confusion Matrix – XceptionNet**

The class activation map for this model was prepared where the activation provided by the last layer is represented as an overlayed heat map. The red-yellow colour depicts the region of interest while the violet-blue colour represents are of low activation

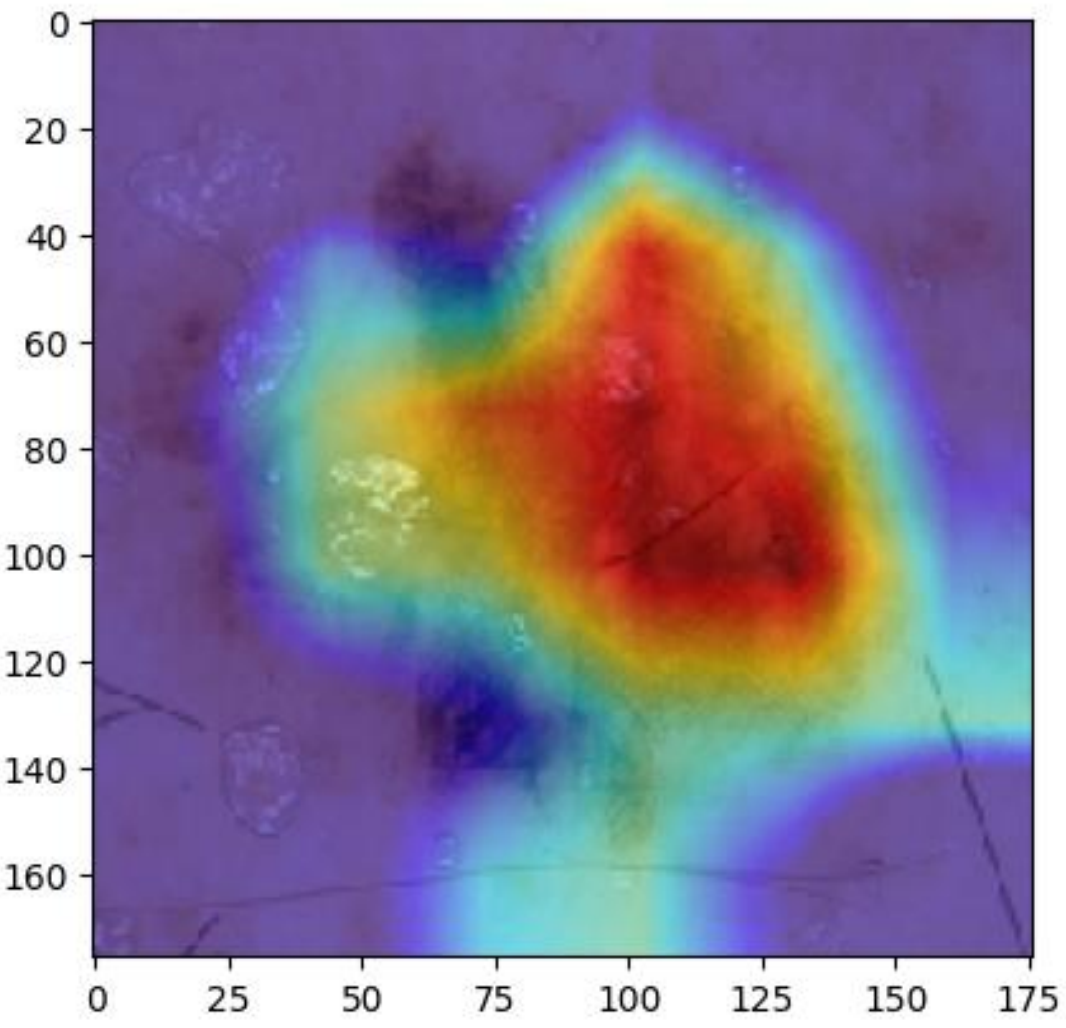

Fig 15: Grad-CAM - XceptionNet

## 3.4 MobileNet

MobileNet is a simpler and lighter model, usually trained for handheld lower computing powered devices. The contemplation behind the usage of MobileNet was to enable the scanning for malignant or benign lesions using globally accessible devices for example smartphones. This model was introduced basically to bring the power of this model to the common. The model performed very decently considering its more efficiency focussed nature. This architecture too was trained over the entirety of the dataset. The model gave an accuracy of over 80.8% as documented in the tabulated data.

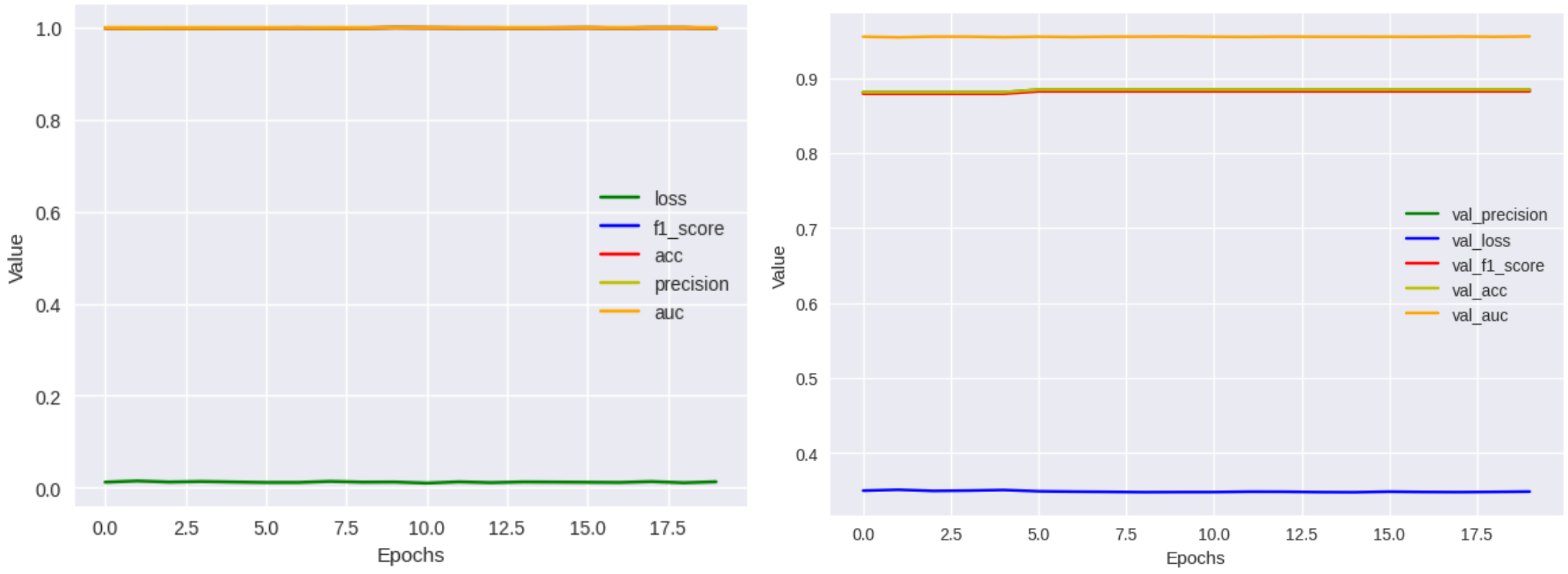

**Fig 16: Graphical Representation of Training and Validation Metrics - MobileNet**

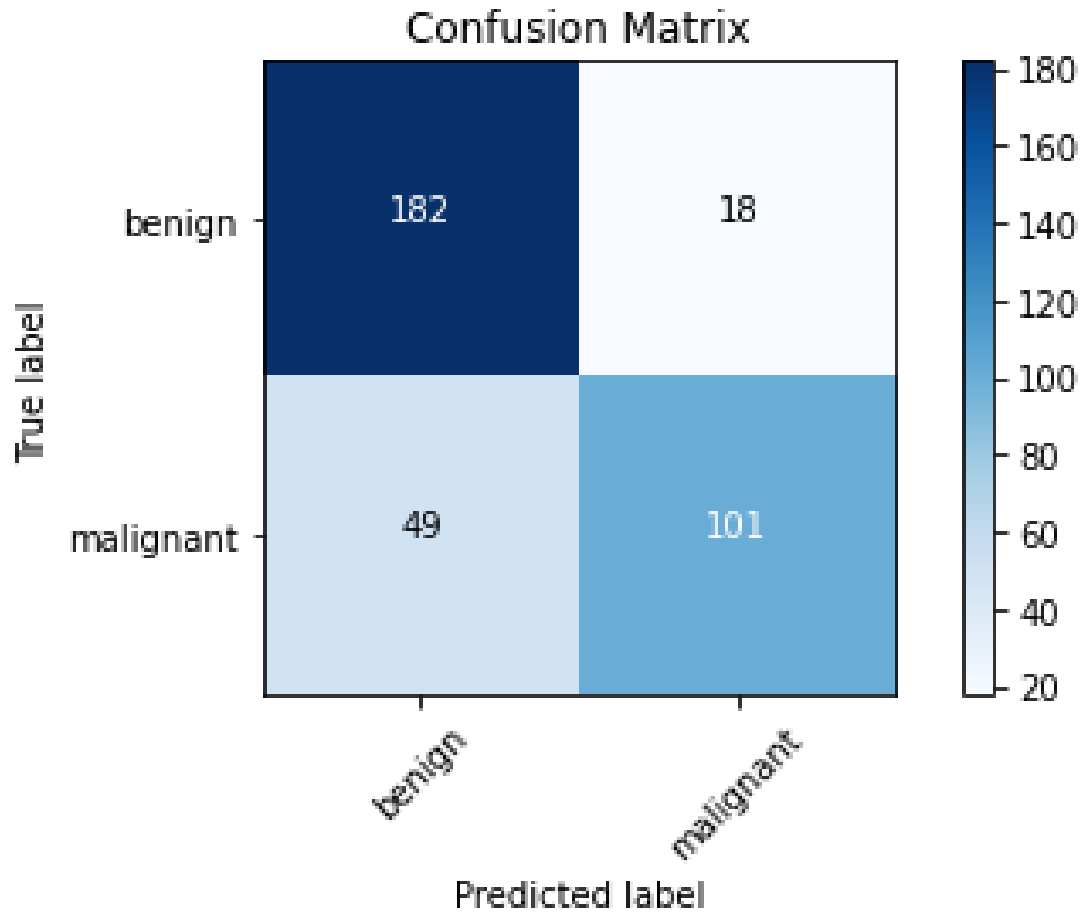

**Fig 17: Confusion Matrix – MobileNet**

The class activation map for this model was prepared where the activation provided by the last layer is represented as an overlayed heat map. The red-yellow colour depicts the region of interest while the violet-blue colour represents are of low activation

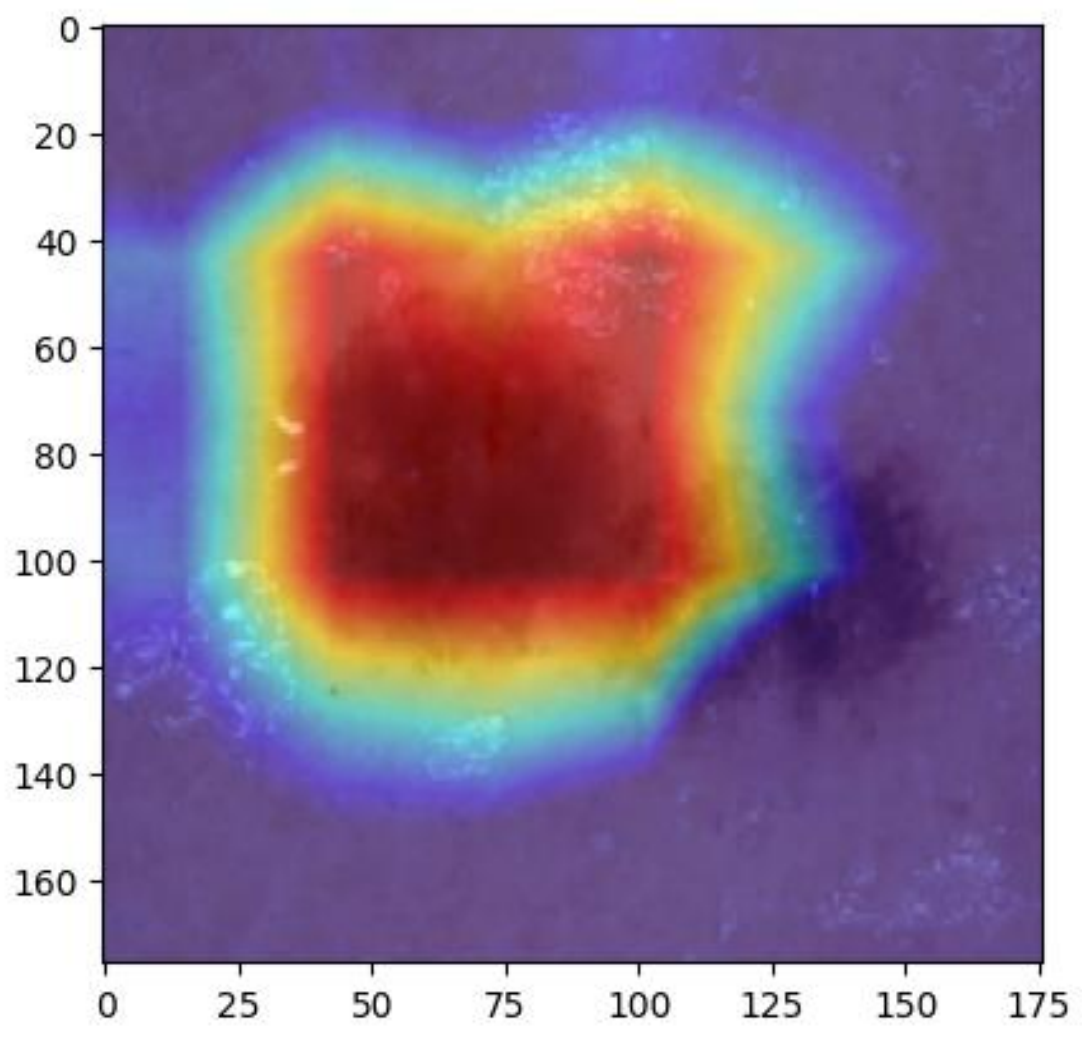

**Fig 18: Grad-CAM – MobileNet**

## 3.5 Comparisons

The results from all the models were compiled and the metrics plotted. Some metrics were tabulated based on the results obtained from the prediction of the test set. The metrics Accuracy, Precision, Recall, and F1-Score were used to compare the model performance among all the chosen architectures.

| | DenseNet | Resnet | XceptionNet | MobileNet |
|---|---|---|---|---|
| Accuracy | 0.86 | 0.86571 | 0.82571 | 0.80857 |
| Precision | 0.8566 | 0.8648 | 0.82555 | 0.81309 |
| Recall | 0.85833 | 0.86 | 0.73991 | 0.79166 |
| F1-Score | 0.85746 | 0.86239 | 0.7803 | 0.802232 |

**Fig 19: Comparison of results for the models used**

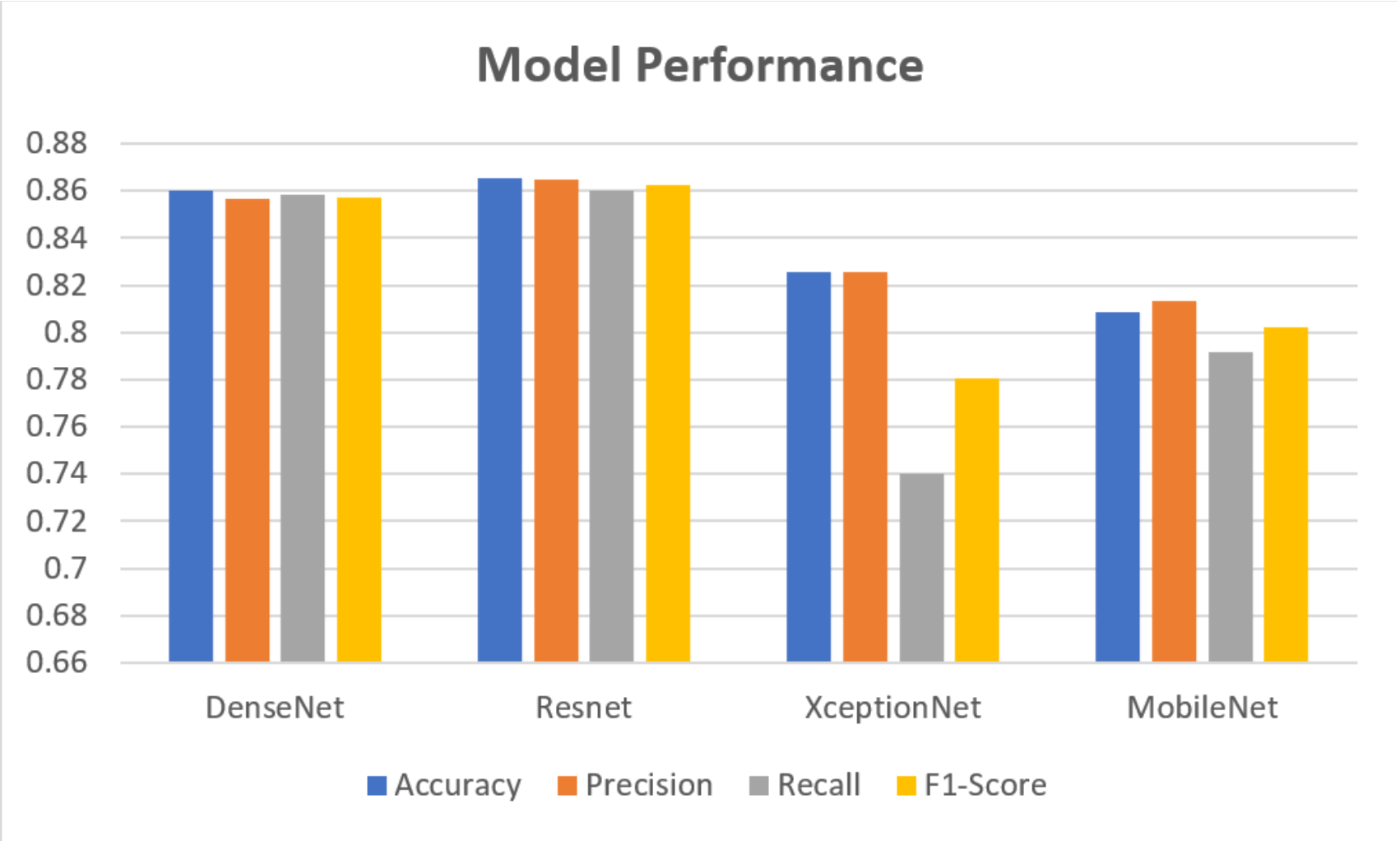

**Fig 20: Graphical Representation of overall results**

## 4 Conclusion and Outlook

Machine Learning and Artificial Intelligence based classification can help in instantaneous diagnosis and classification of Skin Cancer and like diseases. The models developed by us are proof of the concept that cost-effective, user-friendly, and non-invasive Machine Learning and Artificial Intelligence based methods can be developed for the first step towards battling against Skin Cancer and other diseases. During the research process, each model we trained was a successor to the last one, every new model we created was based on the faults of the previous one. All the insights gained from the precursing models were employed to improve the accuracy for the upcoming new models.

In the study presented, the final models were trained and analysed to improve the efficiency and reliability of the methods used for the detection of Skin Cancer. The comparative analysis between various models helped in determining the best model which can have real world applications in the early detection of the disease. The models used in the study presented different fronts in detecting the disease, with each model equipped with better techniques and features. The results obtained from this study have been presented both in written and visual form (vide the help of graphs and confusion matrices) for the perusal of the scientific community.

Further we hope that our findings will help with the early detection of Cancer and be a useful contribution in fighting against this problem. However, human judgement is indispensable. Hence with our model, we are not trying to replace a Dermatologist but merely aid with the predictions.

## 4.1 Future Scope

This study can be understood and referred to as the basis for further analysis in the future. The elementary results explored here may just be the beginning of an exploratory analysis of the power of Machine Learning in the field of Cancer study. Such studies may one day help the general to discover the presence of any malignant tumour easily and get treated in time.